\documentclass{elsart}
\usepackage{graphicx}
\usepackage{graphics}
\usepackage{amssymb}

\begin{document}

\begin{frontmatter}

\title{Fluctuations in time intervals of financial data 
from the view point of the Gini index}

\author[label1]{Naoya Sazuka}
\ead{Naoya.Sazuka@sony.co.jp}
\address[label1]{Sony Corporation, 
4-10-18 Takanawa Minato-ku, 
Tokyo 108-0074, Japan}

\author[label2]{Jun-ichi Inoue}
\ead{j$\underline{\,\,\,}$inoue@complex.eng.hokudai.ac.jp}
\address[label2]{Complex Systems Engineering, 
Graduate School of Information 
Science and Technology, 
Hokkaido University, 
N14-W9, Kita-ku, Sapporo 060-0814, Japan}

\begin{abstract}
We propose an approach 
to explain fluctuations in time intervals 
of financial markets data
from the view point of the Gini index.
We show the explicit form of the
Gini index for a Weibull distribution
which is a good candidate to describe 
the first passage time of
foreign exchange rate. 
The analytical expression 
of the Gini index 
gives a very close value 
with that of empirical data analysis.
\end{abstract}

\end{frontmatter}

\section{Introduction}
\label{sec:Intro}
Almost $10$ years, financial data have attracted a lot 
of attentions of physicists as informative materials to 
investigate the macroscopic behavior of the markets from 
the microscopic statistical properties \cite{Mantegna2000,Bouchaud,Voit}. 
Some of these studies are restricted to 
the stochastic variables of the price changes 
(returns) and most of them is specified by a key word, that is 
to say, {\it fat tails} of the 
distributions \cite{Mantegna2000}. However, 
the distribution of time intervals 
also might have important information about the 
markets and it is worth while for us to investigate these 
properties extensively 
\cite{Engle,Raberto,Scalas,Scalas1,Kaizoji}.

In our previous studies 
\cite{Sazuka,Sazuka2,InoueSazuka2006}, 
we showed that a Weibull distribution 
is a good candidate to describe 
the time intervals of 
the first passage process of foreign exchange rate. 
However, from the shape of the Weibull distribution, 
intuitively, it is not easy to understand 
fluctuations in time intervals. 
To overcome this point, 
in this paper, we introduce a Gini index, 
which is often used in economics to measure 
an inequality of income distribution. 
We here introduce the Gini index 
as a measure of an inequality of the time interval lengths. 
We first derive the Lorentz curve and 
the explicit form of 
the corresponding Gini index 
for a Weibull distribution analytically. 
We show the analytical expression 
of the Gini index is 
in a good agreement with empirical data analysis. 
Then, our analysis makes it possible to 
explain fluctuations in time intervals 
from the view point of the Gini index. 

The paper is organized as follows. 
In the next section, 
we introduce a Gini index and derive 
the analytical expression of the Gini index 
for the Weibull distribution. 
We also evaluate the Gini index for empirical data 
and find a good agreement with empirical data analysis. 
The last section is conclusion and discussions.

\section{Gini index for a Weibull distribution}
\label{sec:Gini}

In our previous studies 
\cite{Sazuka,Sazuka2,InoueSazuka2006}, 
we showed that the distribution of
the time interval between price changes
of the Sony bank USD/JPY rate 
is approximated by a Weibull distribution. 
The Sony bank rate is 
that the Sony bank \cite{Sony} 
offers to their individual customers 
on their online foreign exchange trading service 
via the internet.  
The Sony bank rate is
a kind of first passage processes 
\cite{Redner,Kappen}
with above and below 0.1yen 
for the market rate,
and once the market rate exceeds a threshold, 
the process stops and 
restarts at the updated Sony bank rate. 
Thus, the mean first passage time
of the Sony bank rate
is $\sim$ 20 minutes~\cite{Sazuka},
which is longer than 
the mean time intervals 
of the market rate ($\sim$ 7 seconds).

In this section, 
we investigate the Lorentz curve 
and the corresponding Gini index 
for a Weibull distribution.
The Weibull distribution is 
described by
\begin{eqnarray}
P_{W}(t) & = &
\frac{m t^{m-1}}{a}
\exp
\left(
-\frac{t^{m}}{a}
\right)\,    
\quad 0 \le t < \infty, a>0, m>0
\label{eq:Weibull}
\end{eqnarray}
where $a$ and $m$ are the scale parameter, 
the shape parameter, respectively.
When $m=1$, a Weibull distribution 
is identical to an exponential distribution. 
\begin{figure}[ht]
\begin{center}
\rotatebox{-90}{\includegraphics[width=7cm]{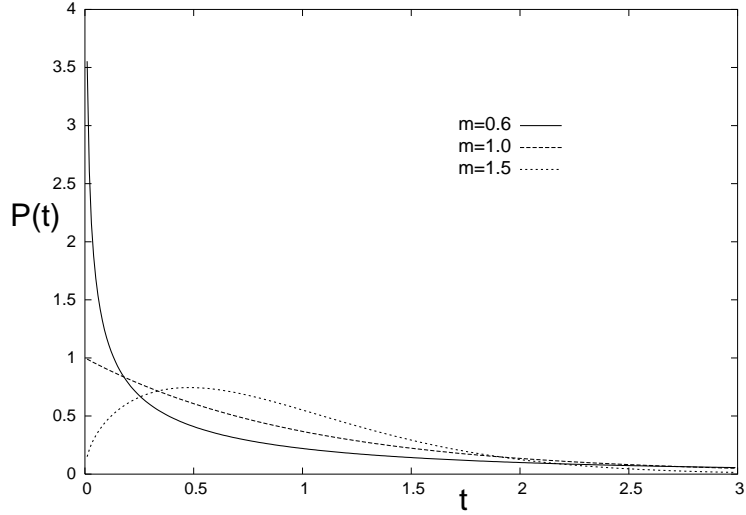}}
\end{center}
\caption{\footnotesize 
Weibull distribution 
for several values of $m$. 
We set a scale parameter $a=1$.}
\label{fig:fg0}
\end{figure}
In FIG. \ref{fig:fg0}, 
we plot the Weibull distribution 
for several values of 
$m$ with $a=1$. 

The Gini index is 
a measure of an inequality in a distribution. 
It is often used in economics 
to measure an inequality of 
income or wealth
in each country or community. 
However, 
we here introduce 
a Gini index 
as a measure of an inequality in 
the length of time interval between data. 
Namely, 
we try to 
recognize the meaning of 
parameter $m$ through the Gini index.
The Gini index 
takes between $0$ 
when all intervals are equal lengths
(perfect equality)
and $1$ when all intervals but one 
are zero lengths
(perfect inequality).

\subsection{Analytical expression of the Gini index}
\label{subsec:Gini}

\begin{figure}[ht]
\begin{center}
\rotatebox{-90}{\includegraphics[width=7cm]{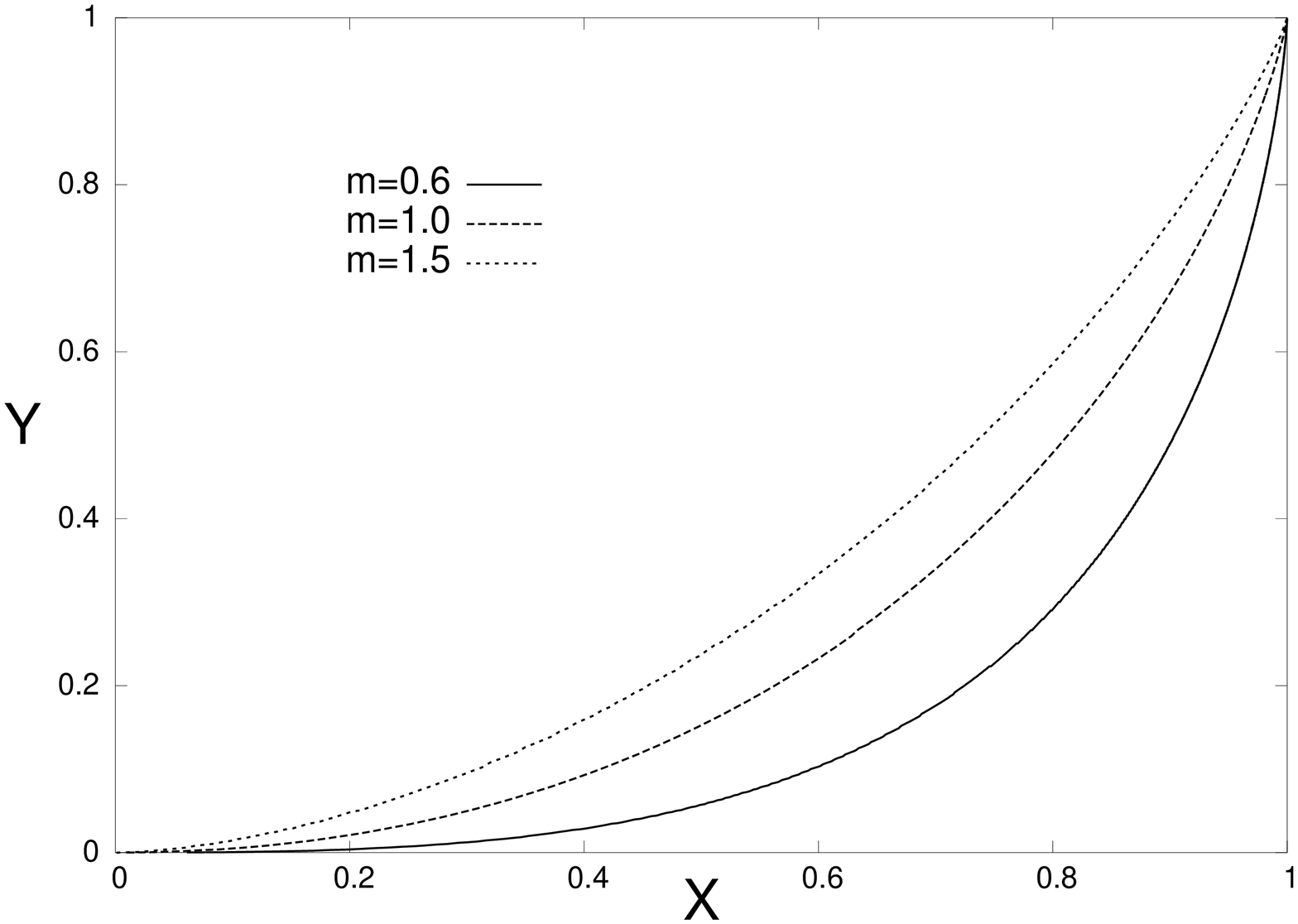}}
\end{center}
\begin{center}
\rotatebox{-90}{\includegraphics[width=7cm]{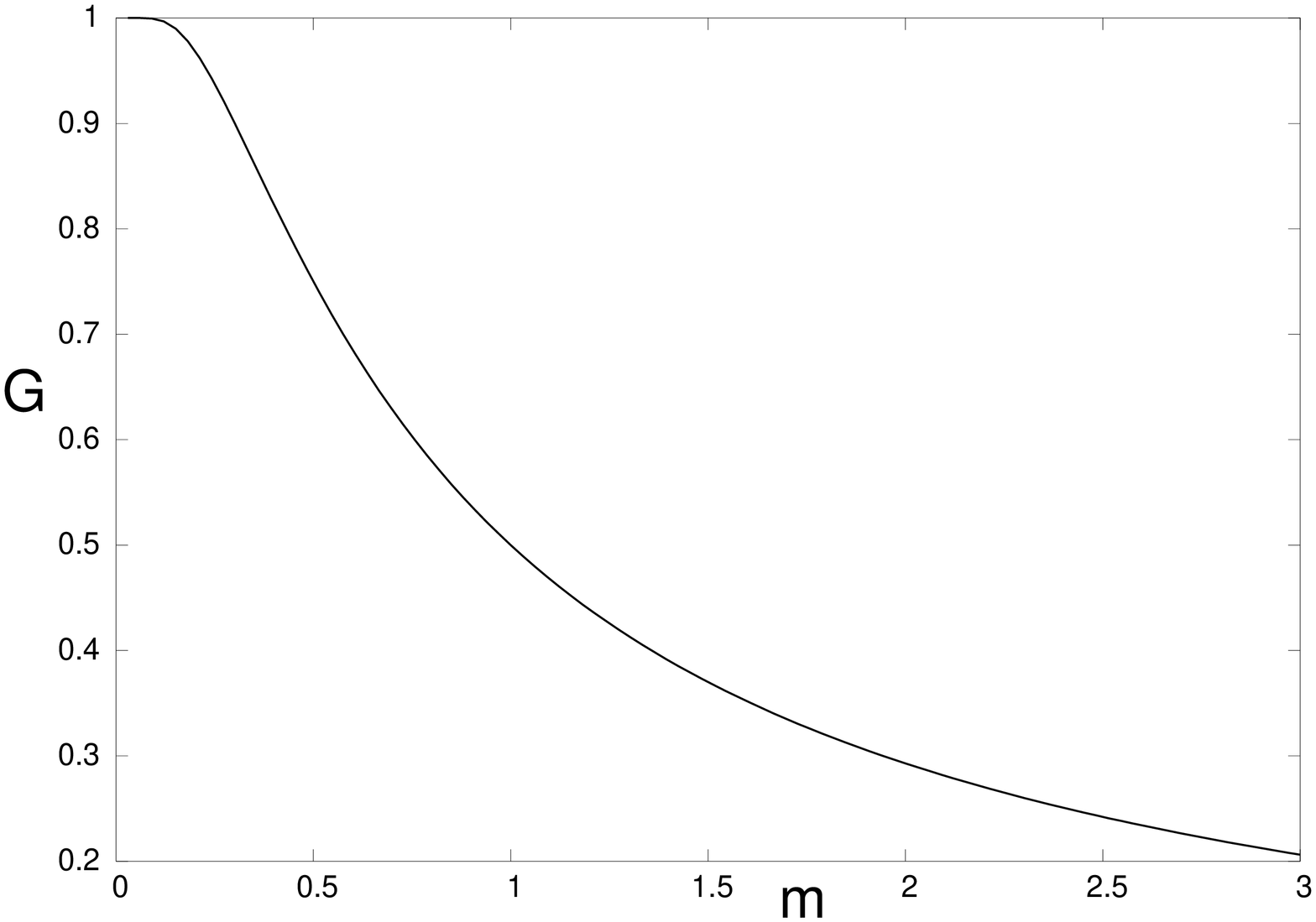}}
\end{center}
\caption{\footnotesize 
The Lorentz curve for a Weibull distribution (upper panel). 
The lower panel shows Gini index 
$G$ as a function of 
$m$ for a Weibull distribution.
}
\label{fig:fg1}
\end{figure}

The Gini index is derived analytically 
if the corresponding ``wealth distribution" is given. 
In this subsection, 
we show explicit form of the Gini index 
for a Weibull distribution as the ``wealth distribution". 
The Gini index 
is derived from the Lorentz curve. 
The Lorentz curve for a 
Weibull distribution is 
described by 
the following relation
between
$
X(r)= 
\int_{0}^{r} 
P_{W}(t) \,dt
$
and
$
Y(r)=
\int_{0}^{r} 
t P_{W}(t) \,dt
/
\int_{0}^{\infty} 
t P_{W}(t) \,dt.
$
Thus, we have 
the Lorentz curve 
for a Weibull distribution as follows. 
\begin{eqnarray}
Y & = &
Q
\left(
\frac{1}{m}+1, 
-\log (1-X)
\right)
\end{eqnarray}
where 
$Q(z,x)$ is the incomplete Gamma function 
given by 
$Q(z,x) = 
\int_{0}^{x}
t^{z-1}
{\rm e}^{-t} dt$. 
In FIG. \ref{fig:fg1} (upper panel), 
we plot the 
Lorentz curve 
for several values of $m$. 
\mbox{}

We next calculate the Gini index 
that is given as twice an area between 
the perfect equality line $Y=X$ and 
the Lorentz curve. 
Carrying out some 
simple algebra,
we have the Gini index 
$G$ as follows. 
\begin{eqnarray}
G & = & 
2 \int_{0}^{1} 
(
X-Y 
) dX
=1-
\left(
\frac{1}{2}
\right)^{1/m}.
\label{eq:G_con}
\end{eqnarray}
It should be noticed that 
the Gini index 
$G$ for a Weibull distribution 
is independent of the 
scale parameter $a$. 
In FIG.  \ref{fig:fg1} (lower panel), 
we plot the Gini index 
$G$ as a function of $m$. 
We find that the Gini index 
$G$ 
monotonically decreases 
as $m$ increases. 
This means that 
for small $m$, 
long intervals are merely 
generated from 
the Weibull distribution, 
whereas, short intervals 
are generated 
with high probability. 
As a result, 
the inequality of 
the interval length 
becomes quite large and 
the Gini index has a value close to $1$. 
For large $m$, on the other hand, 
similar interval lengths
are generated 
from the Weibull distribution.
As a result, 
the inequality is small and
the Gini index is 
close to zero. 
Therefore, 
now the shape of 
the Weibull distribution 
was explained from the 
view point of the inequality of interval length, 
namely, the Gini index. 
As a special case, 
substituting $m=1$ 
into (\ref{eq:G_con})
we can check the Gini index is 
$G=0.5$ for exponential distribution,
which is caused by
the Poisson arrival process of price changes.

Since the empirical value of $m$ 
is about $0.585$
for our data set,
which is around 31,000 data
from September 2002
to May 2004,
the analytical expression of 
the Gini index 
gives $G=0.694168$,
which means more variations than 
the Poisson arrival process
of price changes. 
Therefore, the Sony bank rate has
mainly short intervals and few long intervals.
\subsection{Gini index for empirical data}
\label{subsec:Gini2}
For comparison, 
we next derive 
the Gini index 
for empirical data, 
that is, 
the Gini index 
for discrete probabilistic variables \cite{Xu}. 
Given a sample of $N$ intervals 
with the length $x_i$ 
in non-decreasing order 
$(x_1\le
x_2\le
\cdots
\le x_N)$. 
Discrete probabilistic variables
$X_i$ and $Y_i$, which are ingredients of 
the Lorenz curve,  
are given by 
$X_i=i/N
$
and
$Y_i=
\sum_{r=1}^{i} x_r
/
\sum_{r=1}^{N} x_r
=(\mu N)^{-1}
\sum_{r=1}^{i} x_r
$
for $i=1,2,\cdots,N$. 
A parameter $\mu$ denotes the mean length
$\mu=N^{-1}
\sum_{r=1}^{i} x_r
$ and $X_0$ and $Y_0$ are set to zero.
Thus,
the Gini index for 
the discrete empirical data 
can be obtained as follows.
\begin{eqnarray}
G & = &
\frac{1}{N^2\mu}
\sum_{i=1}^{N}
(2i-N-1) x_i\,.
\label{eq:G_dis}
\end{eqnarray}
From (\ref{eq:G_dis})
the empirical result 
of the Gini index
for the Sony bank rate 
is $G=0.693079$,
which is very close to 
the Gini index 
for the estimated 
Weibull distribution
$G=0.694168$
from (\ref{eq:G_con}).
We find that
the Weibull distribution is 
a plausible candidate
for the time interval distribution 
of the Sony bank rate in terms of the Gini index.
The detail calculations of the Gini index 
will be reported in our forthcoming paper 
\cite{SazukaInoue2007}.
\section{Conclusion and dicussions}
\label{sec:Con}

In this paper,
we proposed an approach 
to explain fluctuations in time intervals 
of financial markets data
from the view point of the Gini index.
We showed the explicit form of the
Gini index for a Weibull distribution
which is a good candidate to describe 
the first passage time of
foreign exchange rate. 
The analytical expression 
gave the very close value
to the empirical data analysis.
More precisely, 
we previously found that
the tails of the time interval distribution 
changes its shape 
from Weibull-law to power-law
\cite{Sazuka2}.
However,
even if without the correction of 
the power-law tail, 
the Gini index 
for a Weibull distribution
is in a good agreement with 
the empirical result.
It is reasonably expected that
the tails of the distribution
does not have a significant effect 
on the Gini index.
Finally,
our approach can be applicable to
other stochastic processes 
to explain fluctuations in intervals.

\section*{Acknowledgement}
One of the authors (N.S.) 
would like to appreciate Shigeru Ishi, 
President of the Sony bank, 
for kindly providing the Sony bank data 
and useful discussions.

\bibliography{apssamp}

\end{document}